\documentclass[conference]{IEEEtran}
\IEEEoverridecommandlockouts

\usepackage{cite}
\usepackage{amsmath,amssymb,amsfonts}
\usepackage{algorithmic}
\usepackage{graphicx}
\usepackage{textcomp}
\usepackage[hyphens]{url}

\usepackage{subcaption}
\usepackage{color}
\usepackage[dvipsnames,svgnames,table]{xcolor}
\usepackage{multirow}
\usepackage{lipsum}
\usepackage{caption}
\usepackage{wrapfig}
\usepackage{tabularx}

\usepackage{makecell}
\usepackage{xspace}
\usepackage{pifont}

\usepackage{balance}

\usepackage[noend, ruled, vlined, linesnumbered, noline, boxed]{algorithm2e}

\usepackage{caption}
\usepackage{booktabs}
\usepackage{etoolbox}

\newcommand{\ceil}[1]{\left\lceil #1 \right\rceil}

\usepackage{lipsum}
\usepackage{adjustbox}
\usepackage{bm}
\usepackage{soul}
\newcommand{\minisection}[1]{\vspace{0.05in}\noindent {\bf #1}}

\newcommand{\name}{\textit{Lit Silicon}{}\xspace}

\usepackage[most]{tcolorbox}
\usepackage{etoolbox} %

\usepackage{tikz}
\DeclareRobustCommand{\circled}[1]{%
  \tikz[baseline=(char.base)]{
    \node[shape=circle,draw=black,fill=white,inner sep=1pt, line width=0.75pt] (char)
    {\color{black}\sffamily\scriptsize #1};}}

\usepackage{hyperref}

\usepackage{threeparttable}

\usepackage{fontawesome}
\newcounter{insightcounter}
\newcommand{\insight}[2][]{
  \refstepcounter{insightcounter}
  \begin{tcolorbox}[
    colframe=gray,
    colback=lightgray,
    boxrule=0.5pt,
    arc=3pt,
    left=0pt,
    right=0pt,
    top=0pt,
    bottom=0pt
  ]
  \ifx\\#1\\%
    \textbf{\faLightbulbO\ Insight \arabic{insightcounter}}: #2
  \else
    \hypertarget{#1}{}%
    \label{#1}%
    \textbf{\faLightbulbO\ {Insight \ref*{#1}}}: #2
  \fi
  \end{tcolorbox}
}

\newcommand{\thetitle}{\name{}: A Case Where Thermal Imbalance Couples Concurrent Execution in Multiple GPUs}
\DeclareMathOperator*{\agg}{agg}
\DeclareMathOperator*{\med}{med}

\def\BibTeX{{\rm B\kern-.05em{\sc i\kern-.025em b}\kern-.08em
    T\kern-.1667em\lower.7ex\hbox{E}\kern-.125emX}}
\begin{document}

\pdfpagewidth=8.5in
\pdfpageheight=11in

\newcommand{\iscasubmissionnumber}{406}

\pagenumbering{arabic}

\title{\thetitle}

\author{
	\IEEEauthorblockN{Marco Kurzynski}
	\IEEEauthorblockA{
		\textit{University of Central Florida}\\
		Orlando, Florida \\
		marco.kurzynski@ucf.edu}
	\and
	\IEEEauthorblockN{Shaizeen Aga}
	\IEEEauthorblockA{
		\textit{Advanced Micro Devices, Inc.}\\
		Santa Clara, California \\
		shaizeen.aga@amd.com}
	\and
	\IEEEauthorblockN{Di Wu}
	\IEEEauthorblockA{
		\textit{University of Central Florida}\\
		Orlando, Florida \\
		di.wu@ucf.edu}
}

\maketitle
\thispagestyle{plain}
\pagestyle{plain}

\begin{abstract}
	GPU systems are increasingly powering modern datacenters at scale.
	Despite being highly performant, GPU systems can exhibit performance variation at the node and cluster levels.
	Such performance variation can significantly impact both high-performance computing and artificial intelligence workloads, such as cutting-edge large language models (LLMs).
	In this work, we analyze the performance of a single-node multi-GPU system running LLM training, and observe that the kernel-level performance variation is highly correlated with concurrent computation and communication (C3), a technique to overlap computation and communication across GPUs for performance gains.
	We then take a further step to reason that thermally induced straggling coupled with C3 impacts performance variation, which we coin the \name effect.
	More specifically, \name describes that in a multi-GPU node, thermal imbalance across GPUs can introduce node-level straggler GPUs (hotter and slower), which in turn slow down the leader GPUs (cooler and faster).
	\name{} can lead to node-level performance variation and inefficiency, potentially impacting the entire datacenter.

	We propose analytical performance and power models for \name, to understand the potential system-level gains.
	We further design simple detection and mitigation techniques to effectively address the \name problem, and evaluate three different power management solutions, including (1) power optimization under GPU thermal design power, (2) performance optimization under node-level GPU power capping, and (3) performance optimization under node-level CPU power sloshing.
	We conduct experiments on two workloads on two AMD Instinct\texttrademark{} MI300X GPU systems under two LLM training frameworks, and observe up to $6\%$ performance and $4\%$ power improvements, potentially saving several tens of millions of dollars in electricity costs in datacenters.
	Our solution consists of approximately 200 lines of PyTorch code, requires no GPU kernel modifications, and can be deployed in datacenters as a new node-level power management layer.
	Our code is available on GitHub: \url{https://github.com/UnaryLab/lit_silicon_tuning_amd}.

\end{abstract}

\section{Introduction}

Due to massively parallel computing capability, GPU systems are gaining wider adoption in modern datacenters to handle compute intensive workloads, either traditional high-performance computing (HPC) workloads (database~\cite{gpu_db_1, gpu_db_2}, scientific computing~\cite{gpu_sc_1, gpu_sc_2}, etc.), or emerging artificial intelligence (AI) workloads (recommendation systems~\cite{gpu_recsys_1, gpu_recsys_2}, content generation~\cite{grattafiori2024llama3herdmodels, gpt3_paper}, etc.).
For such workloads, data transfer easily becomes the system performance bottleneck, due to the large data volume.
To maximize the performance, concurrent computation and communication (C3), a technique that overlaps the computation and communication to hide the communication latency, has been adopted pervasively~\cite{overlap_hpc_1, overlap_hpc_2, overlap_hpc_3}.
C3 has become an indispensable technique to deliver high performance and efficiency in recent AI workloads, such as large language models (LLMs) with billions or trillions of weights~\cite{gpt3_paper, grattafiori2024llama3herdmodels, jiang2024mixtral}, with average speedup between $1.1\times$ and $1.6\times$~\cite{overlap_characterization,ConCCL}.
Such large sizes necessitate sharding models across multiple GPUs, introducing frequent GPU-GPU communication to synchronize model weights, activations, gradients and hyperparameters~\cite{ConCCL, overlap_ai_1, overlap_ai_2}.
Despite end-to-end speedup, it is reported that C3 could impact GPU kernel runtime by an average of $18.9\%$ and up to $40.0\%$~\cite{overlap_characterization}.

\begin{figure}[!t]
	\centering
	\includegraphics[width=\columnwidth]{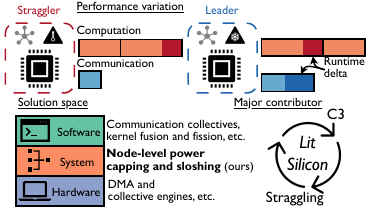}
	\caption{
		Overview of this paper.
		We start from the performance variation in a multi-GPU training, identify the \name effect as a major contributor, and propose solutions to address this effect.
	}
	\label{fig:overview}
\end{figure}

\minisection{Problem.}
There exist diverse parallel strategies to shard LLMs across GPUs, such as data parallel~\cite{dp_paper}, pipeline parallel~\cite{huang2019gpipe, pipedream}, tensor parallel~\cite{tp_paper}, context parallel~\cite{cp_paper}, and expert parallel~\cite{moe_grouped_gemm}.
At the node level, these parallel strategies usually split the full workloads \textit{evenly} across GPUs, and GPU communication is done via collectives over high-bandwidth interconnects~\cite{infinity_fabric_paper, nvlink_paper, nccl_paper, ConCCL}.
For example, during LLM training, fully sharded data parallel (FSDP) shards model weights, activations, and gradients evenly for each layer, and uses communication collectives to synchronize the data~\cite{fsdp_paper}.
FSDP is an identical workload, since each GPU executes operators in the same order with the same dimensions.
\textit{However, even under identical workloads, GPUs in the same node still exhibit strong performance variation in terms of kernel runtime and C3}, as shown in Figure~\ref{fig:overview}.
Such variation separates GPUs in the same node into two groups, slower straggler GPUs and faster leader GPUs, lowering both performance and efficiency.

\minisection{Challenge.}
Knowing the existence of such performance variation and straggling, diverse solutions have been proposed to improve the performance, as shown in Figure~\ref{fig:overview}.
Hardware solutions are usually transparent.
Dedicated direct memory access (DMA) hardware has been extended to ensure better overlapping between computation and communication~\cite{dma_gpu_paper_1, dma_gpu_paper_2}.
There also exists dedicated hardware accelerators for communication collectives~\cite{dma_gpu_acc_paper}.
Software solutions are more fine-grained.
Optimized communication collectives are designed to better hide the latency~\cite{ConCCL, nccl_paper_1, nccl_paper}.
Kernel fusion is used to overlap layer normalization with communication for latency reduction~\cite{c3_kernal_fusion_paper}.
Kernel fission is also leveraged to minimize the idle time on straggler GPUs~\cite{stragglAR}, which assumes a single straggler GPU in the node.

We argue that \textit{to solve the performance variation at the node level effectively, in the presence of C3 and identical workloads, it is critical to understand how it happens.}
However, to the best of our knowledge, no prior work has observed an interplay between performance variation and C3.
Identifying this interplay equips us to address this performance variation challenge in a holistic manner, without costly redesigning of GPU architecture and kernels.

\minisection{Proposal.}
In this paper, we characterize the performance variation and C3 during LLM training, and observe the strong correlation between them.
Then, we identify a major contributor of performance variation as \textit{thermally induced straggling} coupled with \textit{C3} to create an unexpected, often neglected negative feedback loop which we coin the \name effect:

\begin{enumerate}
	\item Thermal imbalance results in leader and straggler GPUs.
	\item Leaders start communication early, but wait for stragglers to complete while the compute stream proceeds independently on all GPUs.
	\item Waiting for stragglers extends communication of leaders and prolongs C3 which slows down leaders.

\end{enumerate}

\name is a dynamic process which repeats at each training iteration, forming a fundamental bottleneck for GPU workloads in the presence of C3 and identical workloads, without losing generality.
To further understand the upper-bound gain for both performance and power, we formulate analytical models for \name.
To solve the \name problem, we craft simple detection and mitigation techniques by tweaking the power caps of individual GPUs within the node, as shown in Figure~\ref{fig:overview}.
We study three unique use cases at the node level, including (1) power optimization under GPU thermal design power (TDP), (2) performance optimization under node-level GPU power capping, and (3) performance optimization under node-level CPU power sloshing.
Our solution essentially introduces a fine-grained node-level power management layer, orthogonal to GPU-level and cluster-level power management, offering datacenter-level performance and power gains.

The contributions of this paper are summarized below.
\begin{itemize}
	\item To the best of our knowledge, we are the first to identify the \name effect, a negative feedback loop in which thermal imbalance and C3 couple to amplify node-level performance variation under identical workloads.
	\item We formulate analytical models to quantify the potential performance and power gains from \name and propose a solution with detection and mitigation techniques.
	\item We evaluate our solution across different workload, software, and hardware settings, and demonstrate consistent gains with low engineering effort required.
\end{itemize}

The rest of the paper is organized as follows.
Section~\ref{sec:background} reviews the background.
Then, Section~\ref{sec:theory},~\ref{sec:model}, and~\ref{sec:architecture} describe our theory, model, and solution for \name.
Next, Section~\ref{sec:implementation} and~\ref{sec:evaluation} evaluate our solution.
Finally, Section~\ref{sec:discussion} and~\ref{sec:conclusion} discuss and conclude this paper.

\section{Background}%
\label{sec:background}

This section briefly reviews the two essential coupling factors of \name{} (i.e., thermally induced straggling and C3) as well as datacenter-level power management, which outlines the solution space of this paper.

\subsection{Thermally Induced Straggling}
Thermally induced straggling describes the performance inefficiency due to overheating.
TDP defines the upper-bound power constraint for reliable execution.
Under TDP, dynamic voltage and frequency scaling (DVFS) further manages the operating voltage and frequency to ensure reliable execution, boost performance and save energy~\cite{dvfs_paper_2, dvfs_dl_paper}.
If overheating, the performance is reported to be lowered by more than $50\%$ in microbenchmarks due to lowered IO bus frequency and enabling advanced ECC, and between $3\%$ and $4\%$ in macrobenchmarks~\cite{thermal_imbalance_paper}.
We term the cooler and faster GPUs as the \textit{leaders}, and the hotter and slower GPUs as the \textit{stragglers}.
Thermally induced straggling exemplifies how device-level power management via DVFS impacts the node- and cluster-level behaviors, regardless of the workloads~\cite{straggler_rootcause_hpc_paper, dvfs_paper_2, dvfs_dl_paper, dvfs_paper_1, stragglAR}.
In this paper, we are concerned with the node-level thermally induced straggling, which is primarily caused by hardware and software, rather than uneven pipeline stage partitioning and across-batch imbalance in sequence lengths~\cite{straggler_rootcause_ai_paper}.

\subsection{Concurrent Computation and Communication}
\label{sec:bckg_c3}
C3 originates from HPC research, where cluster-level performance can be improved by hiding the execution latency of data transfer with computation~\cite{LogP}.
In GPU systems, it means to overlap the execution of computation kernels and communication kernels (i.e., concurrent execution).
C3 is widely used in distributed LLM training to overlap communication kernels, such as AllReduce (AR), AllGather (AG) and ReduceScatter (RS), with computation kernels, especially general matrix multiply (GEMM)~\cite{tp_paper, ZeRO, fsdp_paper, nanoflow, flux, coconet}.
Recent research has predicted that C3's importance will grow in AI workloads, given increasingly larger model size~\cite{tale_of_two_cs}.

\begin{figure}[!t]
	\centering
	\includegraphics[width=\columnwidth]{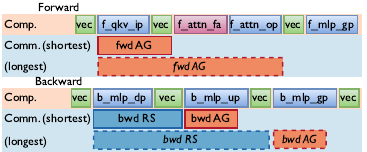}
	\caption{Concurrent computation and communication in FSDP.
		vec: vector operations.
		f\_/b\_: forward/backward.
		qkv\_ip: input projection GEMM of Q/K/V tensors.
		attn: attention.
		fa: flash attention.
		op: output projection GEMM.
		mlp: multi-layer perceptron.
		gp/dp/up: gate/down/up projection GEMM.
	}%
	\label{fig:c3_fsdp}
\end{figure}

We show an example of C3 in an FSDP framework in Figure~\ref{fig:c3_fsdp}, which is based on AG and RS.
In both the forward and backward pass, AG collects shards for the next layer, and in the backward pass, RS reduces gradients for the previous layer.
However, this overlap is not a free lunch, and increases runtime for overlapped kernels.
In the forward phase, AG is at least overlapped with the input projection GEMM of Q/K/V tensors, and reaches as long as the output projection GEMM of the attention layer.
In the backward phase, RS starts to overlap with the down projection GEMM in multi-layer perceptron, and reaches as far as the up projection GEMM.
Then, AG starts to overlap immediately after RS completes.

Traditional manifestation of C3 on GPUs is execution of two concurrent kernels on GPUs (one for compute and one for communication). With finite GPU resources now divvied up among concurrent kernels, C3 suffers from interference from sharing compute and memory resources for concurrent kernels~\cite{ConCCL}, causing undetermined performance variation at the kernel level.
Computation kernels are reported to be slowed down by up to $40\%$~\cite{overlap_characterization}.
This fact makes it very difficult to find the optimal parallelism strategy for GPU systems running AI workloads~\cite{tutel}.
For example, current analytical models to derive the optimal parallelism strategy assume perfect communication collectives with theoretical communication bandwidth and ignore C3 interference~\cite{pipedream}, leading to suboptimal choices.
To mitigate the performance variation from C3, there exist both hardware and software solutions~\cite{dma_gpu_paper_2, ConCCL}.
As an example, communication can be offloaded to DMA engines available on GPUs to reduce compute interference completely and memory interference to some degree~\cite{ConCCL}.
However, such solutions focus on C3 efficiency alone and not performance variation as we aim to tackle in this work.

\subsection{Datacenter Power Oversubscription}%
\label{sec:power_oversub}
Datacenters are built with pre-defined power budget, but can leverage the fact that nodes are usually not fully utilized to add more nodes (i.e., power oversubscription~\cite{power_oversubscription_paper_1}).
Given known workloads, power oversubscription can be done via power capping without significant performance loss~\cite{power_capping_paper_1}.
Power oversubscription has been widely adopted in production environments across industries~\cite{power_oversubscription_paper_1, power_oversubscription_paper_2, power_oversubscription_paper_3, power_oversubscription_paper_4}.
For AI workloads, opportunities for power oversubscription are abundant for inference, and not as rich as for training, since training nearly fully utilizes provisioned power.
However, LLM training suffers from large power swings.
Power capping is an effective means of reducing peak power to limit power swings~\cite{LLMPowerManMicrosoft}.
Therefore, power oversubscription techniques universally exist in datacenters, and we leverage this fact to define the solution space in this paper.
Though we focus on LLM training in this paper, our solution is seamlessly applicable to AI inference.

\section{\name: Characterization}%
\label{sec:theory}

\begin{figure}[!t]
	\centering
	\begin{subfigure}[b]{0.489\textwidth}
		\includegraphics[width=\columnwidth]{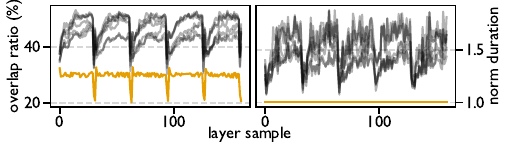}
		\caption{
			Comparison across unique layers.
			Left: the overlap ratio is the weighted average overlap ratio for all kernels in a unique layer, weighted by the computation kernel duration.
			Right: the normalized duration is the sum of all communication kernels in a layer, normalized to the smallest sum across all GPUs.
		}%
		\label{fig:overlap_runtime_corr_layer}
	\end{subfigure}
	\begin{subfigure}[b]{0.489\textwidth}
		\includegraphics[width=\columnwidth]{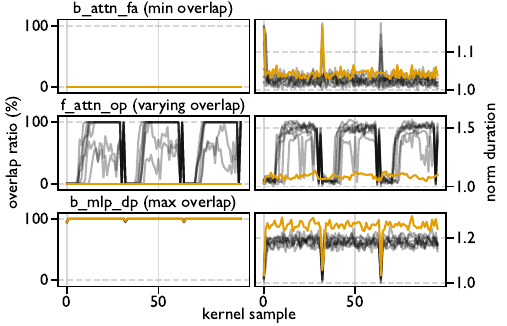}
		\caption{
			Comparison across unique kernels.
			Left: the overlap ratio is the actual overlap ratio of a unique kernel.
			Right: the normalized duration is the actual kernel duration, normalized to the smallest duration across all GPUs\footnotemark.
			b\_attn\_fa and f\_attn\_op are the backward flash attention and forward output projection in attention layer, while the b\_mlp\_dp is the backward down project in multi-layer perceptron.
		}%
		\label{fig:overlap_runtime_corr_op}
	\end{subfigure}
	\caption{
		Comparison between the overlap ratio and the kernel duration for Llama 3.1 8B training over three training iterations.
		Each line represents a unique GPU across time (x-axis), and each sample in a line is for a unique layer or kernel.
		The yellow line marks the straggler GPU, and the gray lines denote the leader GPUs.
		Default settings from Table~\ref{tab:sens_study} are used.
	}%
	\label{fig:overlap_runtime_corr}
\end{figure}
\footnotetext{If an operation includes multiple kernels, the duration counts in the bubbles between these relevant kernels.}

\begin{figure*}[!t]
	\centering
	\includegraphics[width=\textwidth]{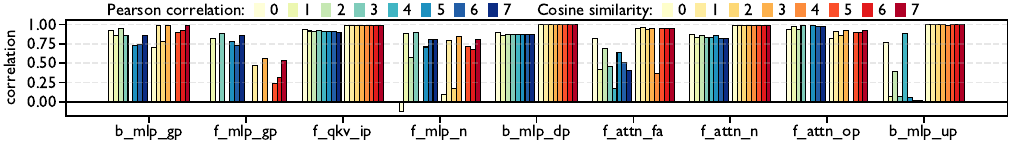}
	\caption{
		Correlation between overlap ratio and kernel duration of kernels across GPUs (numbered).
		f\_/b\_: forward/backward.
		qkv\_ip: input projection GEMM of Q/K/V tensors.
		attn: attention.
		fa: flash attention.
		op: output projection GEMM.
		n: normalization.
		mlp: multi-layer perceptron.
		gp: gate projection GEMM.
		dp: down projection GEMM.
		up: up projection GEMM.
		Default settings from Table~\ref{tab:sens_study} are used.
	}%
	\label{fig:all_op_corr}
\end{figure*}

In this section, we characterize \name by showing the strong correlation between performance variation and C3.
Then, we describe the dynamic process of how straggling accumulates as a result of thermal imbalance across GPUs, and how it couples with C3 to impact performance variation.
Finally, we quantify the potential gains of solving \name by modeling the performance and power.

\subsection{Correlation between Performance Variation and C3}
We profile Llama 3.1 8B training and visualize PyTorch traces with Chopper, a publicly available GPU characterization tool~\cite{chopper} on a node with eight AMD Instinct\texttrademark{} MI300X GPUs under the default training setup from Table~\ref{tab:sens_study} in Section~\ref{sec:implementation}, where all GPUs have identical workloads.
Note that this node has a single straggler GPU.
We show the temporal evolution of the overlap ratio and kernel duration on all GPUs in Figure~\ref{fig:overlap_runtime_corr}.

In Figure~\ref{fig:overlap_runtime_corr_layer}, the overlap ratio and communication kernel duration of all kernels in a unique layer are aggregated and presented.
Here we weight the overlap ratio by the computation kernel duration, to avoid the bias due to shorter but more overlapped kernels, such as vector kernels, as shown in Figure~\ref{fig:c3_fsdp}.
Regarding the overlap ratio, there are four observations.
First, within one iteration, the overlap ratio of all GPUs starts from similar levels, between $30 \%$ and $40 \%$, and the leaders grow their overlap ratio.
Second, within one iteration, the overlap ratio of some leaders reaches a plateau after a few layers, reaching as high as $52.7\%$; others consistently increase the overlap ratio, and do not reach the ratio of plateaued leaders.
Third, the overlap ratio on the straggler GPU remains constant ($29.6\%$) for most of the time and always exhibits the lowest overlap ratio among all GPUs.
The largest overlap ratio of leaders is about $1.8\times$ that of the straggler.
Fourth, across iterations, the overlap ratio pattern almost stays constant for both leaders and straggler, indicating consistent C3 behavior during LLM training.
More importantly, these observations also apply to the communication kernel duration, which intuitively correlates well with the overlap ratio.

\insight[ins:c3-dynamic]{
	Within one training iteration, the straggler GPU has an almost constant C3 pattern.
	The leader GPUs show dynamic C3 patterns, which vary across time and GPUs.
	Across multiple iterations, this dynamic process repeats with a consistent pattern.
}

In Figure~\ref{fig:overlap_runtime_corr_op}, the overlap ratio and communication kernel duration of unique kernels are presented.
We include three iterations of three unique C3 conditions, determined by the overlap ratio.
The first condition is that all GPUs show consistently minimum overlap ratio (e.g., $0\%$ for b\_attn\_fa in the top row).
The second condition is that different GPUs show varying overlap ratio (e.g., between $0\%$ and $100\%$ for f\_attn\_op in the middle row).
The third condition is that all GPUs show consistently maximum overlap ratio (e.g., almost $100\%$ for b\_mlp\_dp in the bottom row).
We define \textit{constant overlap} as kernels with either $0\%$ or $100\%$ overlap on all GPUs, like the first and third condition.
\textit{Varying overlap} is when the overlap is different across GPUs.
Typically, the straggler has the minimum overlap ratio, as shown in the second condition of Figure~\ref{fig:overlap_runtime_corr}.
Again, we observe dynamic and repeated patterns within and across iterations, similar to the findings in Figure~\ref{fig:overlap_runtime_corr_layer}.
And for each operation, there exists a strong correlation between the overlap ratio and kernel duration.
Figure~\ref{fig:all_op_corr} quantifies the degree of correlation between overlap ratio and kernel duration for a few performance-dominant kernels (i.e., GEMM, flash attention, and RMSNorm) using Pearson correlation and cosine similarity.
Both metrics show strong correlation for most kernels and GPUs.

\insight[ins:c3-correlation]{
	The variation in overlap ratio highly correlates with the variation in kernel duration.
	Therefore, C3 has a major impact on across-GPU performance variation in LLM training.
	However, straggler versus leader performance shows contradicting trends under constant versus varying overlap ratio.
}

In addition, we see conflicting behaviors of the straggler versus leaders.
For both the min and max overlap cases (top and bottom), the straggler has higher kernel duration, exhibiting between $5\%$ and $10\%$ lower performance.
On the contrary, for the varying overlap case (middle), the straggler has lower kernel duration, showing $1.5\times$ speedup.
This fact streamlines the formulation of \name as the coupling between thermally induced straggling and C3.

\subsection{Coupling between Thermally Induced Straggling and C3}%
\label{sec:thermal}

\subsubsection{Profiling Thermally Induced Straggling}
\begin{figure}[!t]
	\centering
	\includegraphics[width=\columnwidth]{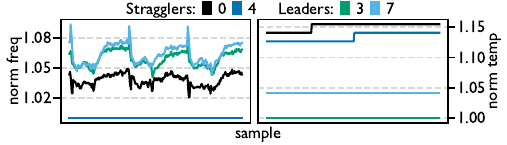}
	\caption{
		Temperature and frequency over three training iterations.
		Both the temperature and frequency are normalized to the lowest value.
		Default settings from Table~\ref{tab:sens_study} are used.
	}%
	\label{fig:temp_freq_corr}
\end{figure}

Figure~\ref{fig:temp_freq_corr} shows the profiled temperature and frequency of two straggler and two leader GPUs, measured with amd-smi~\cite{amdsmi}.
If we take the median of each metric across the samples shown, the highest temperature and frequency are $1.155\times$ and $1.062\times$ those of the lowest values.
Based on the median metric values, if we rank the temperature from high to low for all GPUs, the order is $[0,4,7,3]$, while the order of GPU frequency ranked from low to high is $[4,0,7,3]$.
These two orders are roughly identical, strongly signaling the causality between temperature and frequency across GPUs (i.e., thermally induced straggling).
Despite running the same workload, device-level DVFS is independent of each other, causing variation.
Note that GPU4 in dark blue with the lowest running frequency is not the hottest among all GPUs but the second hottest.
We conjecture that the DVFS management on GPU4 is excessively reducing the frequency when the temperature exceeds a certain level.

\insight[ins:c3-thermal]{
	Within a node, thermal imbalance can induce performance variation across GPUs.
	Higher-temperature, lower frequency stragglers exhibit better performance than leaders for computation kernels with varying overlap ratios (Figure~\ref{fig:overlap_runtime_corr_op}).
}

\begin{figure}[!t]
	\centering
	\includegraphics[width=\columnwidth]{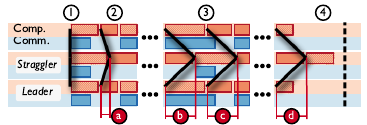}
	\caption{
		Dynamic coupling towards \name.
		\circled{1}-\circled{4} represent four phases of \name in one training iteration.
		The bold black lines, which connect the start time of identical kernels running on different GPUs, are called \textit{straggler waves}.
		The difference in a kernel's start time on a leader and a straggler is defined as the \textit{lead value}.
		\circled{a}-\circled{d} denote the lead values for four different kernels.
	}%
	\label{fig:straggler_wave}
\end{figure}

\subsubsection{Dynamic Coupling towards \name}

Insight~\ref{ins:c3-correlation} and Insight~\ref{ins:c3-thermal} show that both thermally induced straggling and C3 introduce performance variation.
To demonstrate how these factors couple towards \name, we use an example training trace as shown in Figure~\ref{fig:straggler_wave}.
\begin{itemize}
	\item \circled{1} All GPUs start the iteration together.
	      Initially, performance variation is not significant.
	\item \circled{2} The performance variation accumulates across layers.
	      For computation kernels with \textit{constant overlap} (either $0\%$ or $100\%$), leaders run faster and lead values grow.
	      For example, lead value \circled{b} is larger than lead value \circled{a}.

	\item \circled{3} Since the straggler starts communication later, leaders must wait (indicated by the three blue blocks ending together) increasing their overlap.
	      Due to the resource contention during overlap (Section \ref{sec:bckg_c3}), leaders run slower than stragglers for these \textit{varying overlap} kernels.
	      This contention balances out the lead gained from \textit{constant overlap} kernels and \textit{equilibrium} is reached, indicated by identical lead values \circled{b}, \circled{c}, and \circled{d}.
	\item \circled{4} At the end of the iteration, leaders complete all kernels earlier and wait for the straggler to finish.
	      The next iteration will restart the process of \circled{1}-\circled{4}, indicated by the dashed vertical line.
\end{itemize}

\insight[ins:c3-lit]{
	The coupling between thermally induced straggling and C3 has a major impact on performance variation and inefficiency that dynamically accumulates across layers, since communication kernels serve as synchronization points across GPUs.
	The performance variation creates a negative feedback loop, and ultimately balances out to reach equilibrium.
	We coin this dynamic process as \name.
}

\subsection{Degree of Straggling Observed Across Nodes}

\begin{figure}[!t]
	\centering
	\includegraphics[width=\columnwidth]{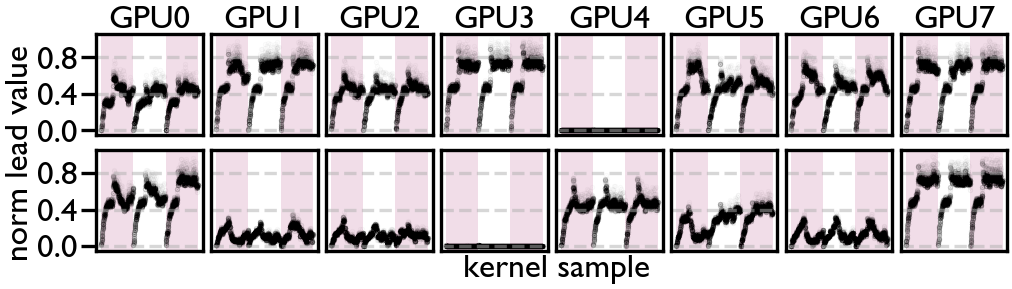}
	\caption{
		Lead values from two test nodes, with node 1 in the top row, and node 0 in the bottom row.
		Each alternating band is for one iteration.
		Default settings from Table~\ref{tab:sens_study} are used.
	}%
	\label{fig:lead_val}
\end{figure}

Knowing the dynamics of \name, we show the profiled lead values from two different training nodes with the same hardware and software configurations in Figure~\ref{fig:lead_val}, and prove \name manifests on both training nodes.
We have a few observations.
First, the patterns of lead values remain almost identical across iteration, indicating that \name is a fundamental issue of such systems.
Second, for the top node, one GPU is absolutely the straggler, since the lead values remain almost constantly at zero.
No other GPUs except GPU4 will have lead values equal to zero, if not at the beginning of an iteration.
Third, for the bottom node, GPUs can take turns being the straggler.
For example, GPU1, GPU2 and GPU6 can now and then become the straggler, though GPU3 claims the straggler position most of the time.
Fourth, the lead values increase on leaders and plateau at certain points, which corroborates the equilibrium.

\section{\name: Modeling Performance and Power}%
\label{sec:model}
\name leads to performance and efficiency loss, and we ask the question: \textit{how much loss does \name introduce?}
Given the dynamic nature of \name, measuring the final wait time at the end of each iteration fails to capture the impact overlap has on leader runtime.
To decompose the dynamics of \name{} into intuitive core concepts, we build analytical models for performance and power.
While these models are not directly used for detection or mitigation of \name{}, the theoretical derivation into Insights~\ref{ins:perf-model}~and~\ref{ins:power-model} helps quantify the key contributor to \name{}: frequency.

\subsection{Performance Model}%
\label{sec:performance_model}
The goal of the performance model is to understand the final performance if we take anti-\name actions that make all GPUs equal (i.e., the same kernels on different GPUs all work identically).
To achieve this, we model the runtime by separating the kernels into two sets based on the overlap ratio, the constant overlap versus varying overlap.
The rationale is that these two kernel sets exhibit the opposite duration trend, as mentioned in Insight~\ref{ins:c3-lit}.
More specifically, leaders are faster for constant overlap kernels, and stragglers are faster for varying overlap kernels.

We first define $\mathcal{G}$ as the set of all GPUs, $\mathcal{K}$ as the set of computation kernels executed on all GPUs.
\begin{equation}
	\mathcal{G} = \{0,\dots,G - 1\},\quad
	\mathcal{K} = \{0,\dots,K - 1\}
\end{equation}
The total runtime can be obtained by summing up the aggregated kernel duration, which are processed from actual profiled traces.
Given $t_{g,k}$ as the kernel duration $k$ executing on GPU $g \in \mathcal{G}$, the total runtime of a set of kernels $t_{\text{agg}}(\mathcal{X})$ is
\begin{equation}
	t_{\text{agg}}(\mathcal{X}) = \sum_{k \in \mathcal{X}}\agg{(t_{\mathcal{G},k})},\ \
	\agg =
	\begin{cases}
		\max \\
		\med \\
		\min
	\end{cases}
	\label{eq:t_agg}
\end{equation}
Here, $\mathcal{X} \in \{\mathcal{C}, \mathcal{V}\}$, where $\mathcal{C} \cup \mathcal{V}=\mathcal{K}$, and $\mathcal{C}$ and $\mathcal{V}$ are the sets of kernels with constant and varying overlap on all GPUs.
The aggregation means we choose the maximum, median, or minimum duration across all GPUs for that kernel.

Therefore, the baseline runtime, confined by the straggler, is given as
\begin{equation}
	t_{\text{baseline}} = t_{\text{max}}(\mathcal{C}) + t_{\text{min}}(\mathcal{V})
	\label{eq:runtime_sys}
\end{equation}
Here, $t_{{\max}}(\mathcal{C})$ is the total runtime of all constant overlap kernels, which have the longest duration on the straggler;
$t_{{\min}}(\mathcal{V})$ is the total runtime of all varying overlap kernels, which have the shortest duration on the straggler.

Starting from the straggler baseline, we can either maintain the runtime or reduce it.
Therefore, the speedup for $\mathcal{C}$ and $\mathcal{V}$, $S_{\mathcal{C}}$ and $S_{\mathcal{V}}$ can be formulated as follows.
\begin{equation}
	S_{\mathcal{C}} = \frac{t_{\max}(\mathcal{C})}{t_{\agg}(\mathcal{C})}, \quad
	S_{\mathcal{V}} = \frac{t_{\min}(\mathcal{V})}{t_{\min}(\mathcal{V})} * S_{\mathcal{C}} = S_{\mathcal{C}}
\end{equation}
$S_{\mathcal{C}}$ indicates the impact of thermally induced straggling (i.e., frequency).
It can have the new runtime (denominator) equal to or smaller than the baseline runtime (numerator), if the frequency is maintained or boosted.
$S_{\mathcal{V}}$ needs to consider the impact of both C3 (the first term) and frequency (the second term).
Since the straggler with the least overlap shows the least runtime for $k \in \mathcal{V}$, it is impossible to speed up these kernels by further reducing the overlap, leading to a constant 1 factor in the first term.
The only opportunity is to boost the frequency via $S_{\mathcal{C}}$.

Next, we leverage Amdahl's law to calculate the speedup of all kernels.
The runtime ratio of $\mathcal{C}$ and $\mathcal{V}$ is $R_{\mathcal{C}}$ and $R_{\mathcal{V}}$.
\begin{equation}
	R_{\mathcal{C}} = \frac{t_{\max}(\mathcal{C})}{t_{\text{baseline}}}, \quad
	R_{\mathcal{V}} = \frac{t_{\min}(\mathcal{V})}{t_{\text{baseline}}}
\end{equation}
Applying Amdahl's law, we finally have the iteration level speedup $S_{\text{iter}}$ as below.
Essentially, the performance improvement is solely determined by boosting the frequency.
\begin{equation}
	S_{\text{iter}} = 1 / (\frac{R_{\mathcal{C}}}{S_{\mathcal{C}}} + \frac{R_{\mathcal{V}}}{S_{\mathcal{V}}}) = S_{\mathcal{C}}
\end{equation}

\insight[ins:perf-model]{
	Speeding up slower overlapped kernels on leaders does not address \name, because the straggler is the fastest for varying overlap kernels.
	The performance is only affected by the difference in frequency across GPUs, and aligning GPU frequencies solves \name.
}

\subsection{Power Model}%
\label{sec:power_model}
The goal of the power model is to obtain the power change ratio under identical optimizations as in the performance model.
We start from a comprehensive power model for AI accelerators~\cite{fine_grained_dvfs}, where $\alpha$, $V$, $f$, $T$ means switching activity ratio, voltage, frequency, and temperature.
For details about other parameters, please refer to the original paper.
\begin{align}
	P                 & = P_{\text{active}} + P_{\text{idle}}        \\
	P_{\text{active}} & = \alpha V^2 f                               \\
	P_{\text{idle}}   & = \beta V^2 f + \gamma \Delta T V + \theta V
\end{align}

In this paper, we assume negligible changes in temperature and voltage and simplify the idle power model to the measured idle power.
This assumption is reasonable, since each GPU exhibits very small temperature variation in the Figure~\ref{fig:temp_freq_corr}.
Then, we can fully control $P_{\text{active}}$ by changing the frequency via power capping, and rewrite it with $M=\alpha V^2$:
\begin{equation}
	P_{\text{active}} = M f
\end{equation}

Furthermore, we assume the relationship between runtime and frequency is identical for all GPUs.
\begin{equation}
	f = \frac{\rho}{t}
\end{equation}

To isolate the impact of overlap on runtime, we only calculate power based on $k \in \mathcal{C}$.
Due to high variation in kernel duration, runtime is summed across ``ranks'' $\mathcal{R}=\{0,...,G-1\}$ instead of GPUs, allowing us to minimize the noise of kernel execution on each GPU.
Kernel durations are sorted and assigned to ranks $r \in \mathcal{R}$, such that kernel duration increases monotonically from $r = 0$ to $r = G-1$.
Then, we have the runtime of rank $r$, $t_{r}$, as the sum of all the rank's kernel durations for $k \in \mathcal{C}$, $t_{r,k}$.
\begin{equation}
	t_{r} = \sum_{k \in \mathcal{C}} t_{r,k}
	\label{eq:rank_time}
\end{equation}
Then, we can have the rank power, $P_{r}$, and system power, $P_{\text{sys}}$, being formulated as below.
\begin{equation}
	P_{r} = M \frac{\rho}{t_{r}} + P_{\text{idle}}, \quad
	P_{\text{sys}} = \sum_{r \in \mathcal{R}}{P_{r}}
	\label{eq:gpu_power_sys}
\end{equation}
Next, we can model the change in power consumption with Equation~\ref{eq:gpu_power_sys}, given $t_{\text{agg}}(\mathcal{C})$ from Equation~\ref{eq:t_agg}.
For each rank, $\delta$ is the multiplicative change in runtime needed to align to $t_{\text{agg}}(\mathcal{C})$, and we have the new rank power $P_{r}'$ as follows.
\begin{equation}
	t_{\text{agg}}(\mathcal{C}) = \delta t_{r} = \delta \frac{M \rho}{P_{r} - P_{\text{idle}}}
\end{equation}
\begin{equation}
	P_{r}' = M \frac{\rho}{t_{\text{agg}}(\mathcal{C})} + P_{\text{idle}} = \frac{P_{r} - P_{\text{idle}}}{\delta} + P_{\text{idle}}
	\label{eq:gpu_power_delta}
\end{equation}

For the baseline with all GPUs running at baseline power, we have $P_{r} = P_{\text{baseline}}$, and get the new rank power and system power as in Equation~\ref{eq:gpu_power_sys}.
Finally, we can use Equation~\ref{eq:gpu_power_sys} and~\ref{eq:gpu_power_delta_mod} to calculate the ratio of power change as $P_{\text{sys}}'/P_{\text{sys}}$.
\begin{equation}
	P_{r}' = \frac{P_{\text{baseline}} - P_{\text{idle}}}{\delta} + P_{\text{idle}}, \quad
	P_{\text{sys}}' = \sum_{r \in \mathcal{G}}{P_{r}'}
	\label{eq:gpu_power_delta_mod}
\end{equation}

\insight[ins:power-model]{
	When mitigating \name by aligning the performance to the straggler/leader GPUs, the power decrease/increase is determined by the number of leader/straggler GPUs, as well as the total difference in frequency.
}

\section{Tackling the \name Effect}%
\label{sec:architecture}

Addressing \name requires a low-overhead and accurate mechanism to detect the straggling, and low-overhead strategies to leverage it, namely saving power, improving performance, or both.
We propose to continuously measure and correct straggling via power capping to reach convergence where no \name is present\footnote{Power capping is reported to be more predictable than frequency capping on GPUs, thus providing more precise control in performance tuning~\cite{power_freq_capping_paper}.}.
The final distribution of GPU power caps after convergence shall hold constant for long-running workloads, such as LLM training.
This means our method only incurs a \textit{one-time} profiling cost, after which it can optionally be disabled, or use a long sampling period, without impacting workload execution.
Our solution is lightweight, with only about 200 lines of PyTorch code, and is applicable to different use cases, where both node-level and GPU-level power caps are considered.
Notations will follow those in performance and power modeling in Section~\ref{sec:model}.

\subsection{Framework and Use Cases}

\begin{figure}[!t]
	\centering
	\includegraphics[width=\columnwidth]{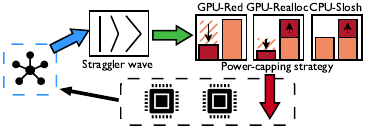}
	\caption{
		Our framework to solve \name with three use cases.
		It only needs about 200 lines of PyTorch code.
	}%
	\label{fig:solution_framework}
\end{figure}

We show the framework of our solution in Figure~\ref{fig:solution_framework}.
Table~\ref{tab:scenarios} outlines three supported use cases, all originating from power oversubscription in datacenters (Section~\ref{sec:power_oversub}).

\minisection{GPU-Red.}
Leaders burn power only to be held back by stragglers during synchronization.
As such, GPU-Red, short for GPU-Reduce, strategically power caps leaders in a dynamic and bespoke manner to realize power savings without losing throughput.

\minisection{GPU-Realloc.}
Stragglers could benefit from boosting power to increase frequency and catch up with leaders, instead of holding them back.
Knowing that leaders burn more power than necessary, we can reallocate the power across GPUs and move the system equilibrium toward superior performance, which is denoted as GPU-Realloc.
Moreover, exceeding TDP at the millisecond level has been standardized~\cite{whitney2019ocp}, where GPU-Realloc can have more room to take effect.

\minisection{CPU-Slosh.}
Finally, we also profile CPU behavior during LLM training, and our profiling results indicate that only $13.5\%$ out of all CPU cores are utilized during training.
This means about $86.5\%$ of the core power, or hundreds of watts, is wasted and could be sloshed to the GPUs to improve performance, which we call CPU-Slosh.
Similar heterogeneous power partitioning has been studied before~\cite{power_partition_cpu_gpu}.

\begin{table}
	\centering
	\caption{
		Use cases of our solution.
	}
	\begin{tabular}{lll}
		\toprule
		\midrule
		\textbf{Use case}            & \textbf{Condition}     & \textbf{Expected outcome} \\
		\midrule
		\multirow{3}{*}{GPU-Red}     & No node-level power    & Node power reduced,       \\
		                             & cap; reduce power      & avg. GPU power reduced,   \\
		                             & on leaders only.       & throughput unchanged.     \\
		\midrule
		\multirow{3}{*}{GPU-Realloc} & Node-level power cap;  & Node power unchanged,     \\
		                             & reallocate power from  & avg. GPU power unchanged, \\
		                             & leaders to stragglers. & throughput increased.     \\
		\midrule
		\multirow{3}{*}{CPU-Slosh}   & Node-level power cap;  & Node power unchanged,     \\
		                             & slosh power budget     & avg. GPU power increased, \\
		                             & from CPU to GPUs.      & throughput increased.     \\
		\midrule
		\bottomrule
	\end{tabular}%
	\label{tab:scenarios}
\end{table}

\subsection{Detection of \name}
\name can be quantified by lead values and detected using a straggler wave in Figure~\ref{fig:straggler_wave}, generated from a trace using Algorithm~\ref{alg:lead_value}.
This algorithm uses the starting timestamp of all kernels across GPUs to calculate the lead values (line~\ref{alg:lead_vals}).
For example, if GPU0 starts a kernel 10ms later than GPU1, then GPU1 has a lead of 10ms for that kernel.
In line~\ref{alg:lead_sum}, we aggregate the lead values for each GPU by summing them up, giving a per GPU lead value vector.
For example, if a GPU's lead increases linearly from 0 to 10ms over 100 kernels, its aggregate lead value would be 500ms.
This per GPU aggregated lead is the output of Algorithm~\ref{alg:lead_value}.
Summing the lead values essentially retrieves the area under the lead value curve in Figure~\ref{eq:t_agg}.
Note that instead of summation, the maximum or the last value of the lead values can be used for aggregation, which will be evaluated later.

\begin{algorithm}[!t]
	\caption{\textsc{LeadValueDetect}}%
	\label{alg:lead_value}
	\KwIn{Timestamp vector $T[g, k]$ for $g \in \mathcal{G}$ and $k \in \mathcal{K}$}
	\KwOut{Lead value vector $L[g]$ for $g \in \mathcal{G}$}

	\ForEach{Kernel $k$}{
		$T_{\textit{max}} \leftarrow \max{(T[\mathcal{G}, k])}$\;
		\ForEach{GPU $g$}{
			$lead\_value[g, k] \leftarrow T_{\textit{max}} - T[g, k]$\;\label{alg:lead_vals}
		}
	}
	\ForEach{GPU $g$}{
		$L[g] \leftarrow \sum\limits_{k}{lead\_value[g, k]}$\;\label{alg:lead_sum}
	}
	\Return $L$\;
\end{algorithm}

\subsection{Mitigation of \name}
\name{} is mitigated using the aggregate lead values from Algorithm~\ref{alg:lead_value} as input to Algorithm~\ref{alg:inc_power_cap} which calculates ideal power-cap increases without TDP or node-level power considered.
Finally, Algorithm~\ref{alg:adj_power_cap} uses the ideal power-caps to uniformly adjust all GPUs to meet the node-level power cap and not exceed TDP.
These algorithms are used for all use cases summarized in Table~\ref{tab:scenarios}, where the only variable that changes per use case is the node-level power cap.
This power cap is decided by the datacenter in production, based on how oversubscribed the datacenter is, and if power-gating idle CPU cores is supported.

To explain how these algorithms apply to each use case, we will use an example of a node with a single straggler, and seven leaders using example values.
Note that the actual parameters and values used are in Table~\ref{tab:sens_study}.

\minisection{GPU-Red.}
The node-level power cap is equal to the maximum provisioned power where all GPUs are running at TDP for the baseline.
Algorithm~\ref{alg:lead_value} detects a single straggler, and Algorithm~\ref{alg:inc_power_cap} requests to increase the straggler's power cap by 15W (the default value for the max adjustment in Table~\ref{tab:sens_study}).
To not exceed TDP, Algorithm~\ref{alg:adj_power_cap} will instead lower the power cap of leaders by 15W and leave the straggler at TDP.

\minisection{GPU-Realloc.}
If the node-level power cap is 120W below the maximum provisioned power, then all GPUs are 15W below the TDP for the baseline.
Algorithm~\ref{alg:inc_power_cap} requests to raise the straggler's power cap by 15W, which would not exceed TDP, but would exceed the node-level power cap.
This time, Algorithm~\ref{alg:adj_power_cap} will increase the straggler's power cap by 15W, then uniformly lower all GPUs by $\frac{15\text{W}}{\text{GPUs}}$.

\minisection{CPU-Slosh.}
The baseline is the same as GPU-Realloc.
The difference is we have a power budget available from the CPU.
If our per GPU power budget is at least 2W, then the straggler's power cap can be increased by 15W without lowering caps on leaders since we have an additional 16W of total power available before reaching the node-level power cap.

The goal of straggler mitigation is to minimize the lead values by tuning the power caps of each GPU.
Theoretically, we can align the distribution of the actual power caps across GPUs towards an expected distribution from the performance and power models.
However, such precise alignment may require long latency to converge.
Therefore, we design Algorithm~\ref{alg:inc_power_cap} and Algorithm~\ref{alg:adj_power_cap} for fast convergence with decent accuracy.

Algorithm~\ref{alg:inc_power_cap} calculates the delta to increase the power cap on each GPU.
It takes in the lead value vector from Algorithm~\ref{alg:lead_value}, a user-defined max increase value of the power cap to avoid over tuning, and the largest lead value across iterations.
The final power cap increase vector of a GPU is proportional to its relative lead values within the current sampled iteration (line~\ref{alg:cur_lead}) and across all past sampled iterations (line ~\ref{alg:past_lead}), which are meant to tune each GPU power separately and ensure the power cap increases are gradually lowered.

\begin{algorithm}[!t]
	\caption{\textsc{IncPowerGPU}}%
	\label{alg:inc_power_cap}
	\KwIn{Lead value vector $L[g]$ for $g \in \mathcal{G}$, maximum value to increase the power cap $max\_inc$, and the largest lead value observed across iterations $global\_max$}
	\KwOut{Power cap increase vector $I[g]$ for $g \in \mathcal{G}$ and updated $\textit{global\_max}$}

	$\textit{max\_lead} \leftarrow \max(L[\mathcal{G}])$\;
	$\textit{min\_lead} \leftarrow \min(L[\mathcal{G}])$\;
	$\textit{global\_max} \leftarrow \max(\textit{global\_max}, max\_lead)$\;
	\ForEach{GPU $g$}{
		$norm\_lead \leftarrow 1 - \frac{L[g] - min\_lead}{max\_lead - min\_lead}$\;\label{alg:cur_lead}
		$I[g] \leftarrow norm\_lead \times \frac{max\_lead}{\textit{global\_max}} \times \textit{max\_inc}$\;\label{alg:past_lead}
	}

	\Return $I, \textit{global\_max}$\;
\end{algorithm}

Algorithm~\ref{alg:adj_power_cap} further tunes the GPU power caps by considering the node-level power cap.
It first increases GPU power caps based on the returned GPU power caps from Algorithm~\ref{alg:inc_power_cap} (line~\ref{alg:output_power_inc}) and updates the total node power (line~\ref{alg:total_node_pow}).
Then, we assume the node-level power increase is uniformly allocated to each GPU and obtain the per-GPU maximum power cap delta (line~\ref{alg:ceil}), which is further adjusted by the GPU TDP to get the actual power cap delta (line~\ref{alg:tdp_power_cap}).
Finally, all GPUs will tune their power cap by the same delta (line~\ref{alg:final_power_cap}).
The output of Algorithm~\ref{alg:inc_power_cap} is the final new power cap of each GPU, and the system sets the power caps accordingly.

\begin{algorithm}[!t]
	\caption{\textsc{AdjPowerNode}}%
	\label{alg:adj_power_cap}
	\KwIn{Power cap increase vector $I[g]$ for $g \in \mathcal{G}$, current power cap vector $P[g]$ for $g \in \mathcal{G}$, maximum power of GPUs $\textit{TDP}$, and node-level power cap $P_n$}
	\KwOut{Updated power cap vector $P'[g]$ for $g \in \mathcal{G}$}

	$\textit{node\_power} = 0$\;
	\ForEach{GPU $g$}{
	$P'[g] \leftarrow P[g] + I[g]$\;\label{alg:output_power_inc}
	$\textit{node\_power} \leftarrow \textit{node\_power} + P'[g]$\;\label{alg:total_node_pow}
	}

	$\textit{gpu\_delta\_max} \leftarrow \ceil{(\textit{node\_power} - P_n) / G}$\;\label{alg:ceil}

	$\textit{gpu\_delta} \leftarrow 0$\;
	\ForEach{GPU $g$}{
	$P'[g] \leftarrow P'[g] - \textit{gpu\_delta\_max}$\;
	$\textit{gpu\_delta} \leftarrow \max(\textit{gpu\_delta}, P'[g] - \textit{TDP})$\;\label{alg:tdp_power_cap}
	}
	\ForEach{GPU $g$}{
	$P'[g] \leftarrow P'[g] - gpu\_delta$\;\label{alg:final_power_cap}
	}
	\Return $P'$\;

\end{algorithm}

\section{Evaluation Setup}%
\label{sec:implementation}

All evaluation knobs are listed in Table~\ref{tab:sens_study}.

\begin{table}[h!]
	\centering
	\begin{threeparttable}
		\caption{
			Evaluation knobs.
		}
		\begin{tabular}{llcc}
			\toprule
			\midrule
			\textbf{Category} & \textbf{Knob}                & \textbf{Values}  & \textbf{Default}              \\
			\midrule
			\textbf{Hardware} & Node                         & 0, 1             & 1                             \\
			\midrule
			\multirow{5}{*}{\textbf{\makecell[l]{Workload                                                       \\ and \\ framework}}}
			                  & Model                        & Llama 3.1 8B,    & \multirow{2}{*}{Llama 3.1 8B} \\
			                  &                              & Mistral 7B v0.1  &                               \\
			\cmidrule(lr){2-4}
			                  & FSDP                         & v1, v2           & v2                            \\
			\cmidrule(lr){2-4}
			                  & Precision\footnotemark[3]    & bf16, fp8        & bf16                          \\
			\midrule
			\multirow{2}{*}{\textbf{Configuration}}
			                  & Batch size,                  & b1s4, b2s4, b4s4 & \multirow{2}{*}{b2s4}         \\
			                  & sequence length              & b1s8, b2s8       &                               \\
			\midrule
			\multirow{3.75}{*}{\textbf{\makecell[l]{Baseline                                                    \\ calibration}}}
			                  & Iterations                   & 1000             & 1000                          \\
			\cmidrule(lr){2-4}
			                  & Sampling period              & 4, 7, 10         & 10                            \\
			\cmidrule(lr){2-4}
			                  & Warm-up                      & 3, 6, 12, 25, 50 & 50                            \\
			\midrule
			\multirow{2.5}{*}{\textbf{\makecell[l]{Straggler                                                    \\ detection}}}
			                  & Window size                  & 1, 2, 3, 5       & 3                             \\
			\cmidrule(lr){2-4}
			                  & Aggregation                  & max, last, sum   & sum                           \\
			\midrule
			\multirow{6}{*}{\textbf{\makecell[l]{Straggler                                                      \\ mitigation}}}
			                  & Max adjustment               & 5, 10, 15, 30    & 15                            \\
			\cmidrule(lr){2-4}
			                  & Scale                        & global, local    & global                        \\
			\cmidrule(lr){2-4}
			                  & Power caps\footnotemark[4]   & 700, 650, 600,   & \multirow{2}{*}{700}          \\
			                  &                              & 550, 500         &                               \\
			\cmidrule(lr){2-4}
			                  & Power budget\footnotemark[5] & 10, 20, 30, 50   & 20                            \\
			\midrule
			\bottomrule
		\end{tabular}%
		\label{tab:sens_study}
		\begin{tablenotes}
			\footnotesize
			\item[3]{FSDPv1 is used for compatibility with Transformer Engine.}
			\item[4]{Only for GPU-Realloc and CPU-Slosh.}
			\item[5]{Only for CPU-Slosh.}
		\end{tablenotes}
	\end{threeparttable}
\end{table}

\minisection{Hardware.}
We use two AMD GPU nodes, each with eight AMD Instinct\texttrademark{} MI300X GPUs and two AMD EPYC\texttrademark{} 9684X CPUs.

\minisection{Workload and framework.}
We evaluate LLM training with FSDP and FSDP2, using two different workloads: Llama 3.1 8B and Mistral 7B v0.1.
FSDP2 improves over FSDP by introducing a new distributed tensor format to better handle the tensor metadata.
Precision is explored by training with bf16 and fp8, where fp8 uses Transformer Engine kernels, with E4M3 for forward (higher precision) and E5M2 for backward (larger range), plus dynamic scaling for stability.

\minisection{Configuration.}
The configurations of batch size and sequence length are chosen that fit in the GPU HBM.
Batch size 2 and sequence length 4k are selected as default, since it is representative in terms of performance and power response to our solution, and also allows collecting traces faster.

\minisection{Baseline calibration.}
Obtaining an accurate baseline is crucial to accurately measure performance and power improvements.
The iteration defines the length of an experiment, and needs to be long enough to reach convergence.
The sampling period defines the interval between sampling an iteration.
Finally, warm-up defines how many samples should be taken before making adjustments to power.

\minisection{Straggler detection.}
The aggregation uses a ``straggler wave'' from Figure~\ref{fig:straggler_wave} to detect stragglers and leaders.
Max takes the largest lead value, last takes the final lead value, and sum is the ``area under the curve'' or sum of lead values for each GPU.
We choose sum as the default for Algorithm~\ref{alg:lead_value} because it still penalizes GPUs while they are in equilibrium.
In theory, this helps to identify leaders in the presence of multiplicative C3 interference.
In practice, max, last, or sum all converge to the expected outcome.
The window size defines how many sample aggregations should be averaged together before adjusting power.

\minisection{Straggler mitigation.}
Max adjustment is the user-defined max power increase value used in Algorithm~\ref{alg:inc_power_cap}.
Using a large max adjustment speeds up convergence at the risk of overshooting stable power caps.
Using a global scale adjusts power less as convergence is approached by tracking the largest lead seen.
A local scale will always use the max adjustment which can speed up convergence at the cost of variation.

\section{Evaluation}%
\label{sec:evaluation}

In this section, we evaluate the benefits and behavior of our straggler detection and mitigation strategies.

\begin{figure}[!t]
	\centering
	\begin{subfigure}[b]{0.489\textwidth}
		\includegraphics[width=\columnwidth]{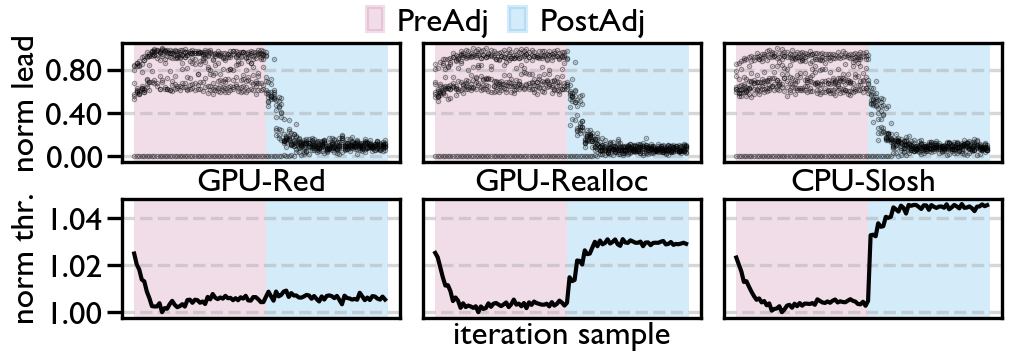}
		\caption{
			Aggregated lead values and throughput of b2s4 for all use cases.
			The aggregated lead value uses summation per GPU.
			Throughput is calculated using the sum of kernel duration.
			The x-axes are sampled iterations.
			The y-axes are normalized to the maximum lead and minimum throughput per use case.
		}%
		\label{fig:lead_and_throughput_all_use_cases}
	\end{subfigure}
	\begin{subfigure}[b]{0.489\textwidth}
		\includegraphics[width=\columnwidth]{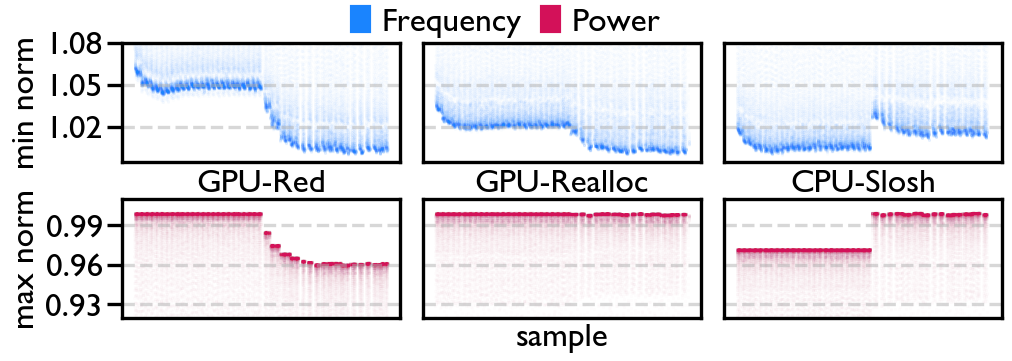}
		\caption{
			Total power of b2s4 for all use cases.
			The x-axes are samples of frequency and power.
			The y-axes are the average frequency and power across GPUs, normalized to the min and max per use case.
			Tuning begins halfway.
		}%
		\label{fig:total_power_and_freq_all_use_cases}
	\end{subfigure}
	\caption{
		Visualization of the convergence process for all use cases using default settings from Table~\ref{tab:sens_study}.
	}%
	\label{fig:all_use_cases_power_throughput}
\end{figure}

\subsection{Overall Comparison across Use Cases}
Figure~\ref{fig:all_use_cases_power_throughput} visualizes each use case dynamically.

\minisection{GPU-Red.}
Reducing power on leaders results in almost no throughput change and reduces lead post adjustment in Figure~\ref{fig:lead_and_throughput_all_use_cases}.
Average power decreases by $4\%$, proportionally to average frequency as shown in Figure~\ref{fig:total_power_and_freq_all_use_cases}.

\minisection{GPU-Realloc.}
Reallocating power to stragglers results in a throughput improvement of $3\%$, and reduces lead in Figure~\ref{fig:lead_and_throughput_all_use_cases}.
This throughput increase is accomplished without raising average power as shown in Figure~\ref{fig:total_power_and_freq_all_use_cases}.
Additionally, the average frequency decreases as a result of allocating more power to thermally inefficient GPUs.

\minisection{CPU-Slosh.}
Sloshing enables reallocating extra power to stragglers, which results in a throughput improvement of $4\%$, and minimizing lead in Figure~\ref{fig:lead_and_throughput_all_use_cases}.
However, this is a result of allocating $3\%$ more power to the GPUs as shown in Figure~\ref{fig:total_power_and_freq_all_use_cases}.

\minisection{Comparison.}
Between the three use cases, GPU-Red provides the greatest benefit of a $4\%$ power reduction. GPU-Realloc increases throughput by $3\%$ without increasing power consumption. Finally, CPU-Slosh marginally improves throughput compared to GPU-Realloc by $4\%$, while consuming $3\%$ more power. The trend is that \textit{allocating more power to stragglers has diminishing returns}.
However, considering the node level power is maintained, this approach also does not increase power consumption in datacenters.

\minisection{Performance and Power Models.}
We compare measured results to predicted results in Table~\ref{tab:model_output} using our performance and power models from Section~\ref{sec:performance_model}~and~\ref{sec:power_model}.
For aligning GPUs with Equation~\ref{eq:t_agg}, we use $\min$, $\med$, and $\max$ as our $\agg$ function for GPU-Red, GPU-Realloc, and CPU-Slosh respectively.
The predicted power is accurate, with $1\%$ error at most.
While the predicted throughput is $2\times$ larger than measured throughput, it captures the trend of diminishing returns of allocating more power to stragglers, going from GPU-Realloc to CPU-Slosh.
Finer-grained modeling by removing some of our assumptions could potentially close the gap.

\minisection{Takeaway.}
The \name effect can be tackled by allocating more power to stragglers, but we see diminishing returns as the amount of power reallocated grows from GPU-Red to GPU-Realloc to CPU-Slosh.

\begin{table}[!h]
	\centering
	\begin{tabular}{lcccc}
		\toprule
		\midrule
		\textbf{Scenario} & \multicolumn{2}{c}{\textbf{Power}} & \multicolumn{2}{c}{\textbf{Throughput}}                        \\
		\cmidrule(lr){2-3} \cmidrule(lr){4-5}
		                  & Predicted                          & Measured                                & Predicted & Measured \\
		\midrule
		GPU-Red           & 1.05                               & 1.04                                    & 1.00      & 1.00     \\
		GPU-Realloc       & 1.00                               & 1.00                                    & 1.06      & 1.03     \\
		CPU-Slosh         & 0.97                               & 0.97                                    & 1.10      & 1.04     \\
		\midrule
		\bottomrule
	\end{tabular}
	\caption{
		Predicted benefit for different use cases using default settings in Table~\ref{tab:sens_study}.
	}%
	\label{tab:model_output}
\end{table}

\begin{figure}[!t]
	\centering
	\includegraphics[width=\columnwidth]{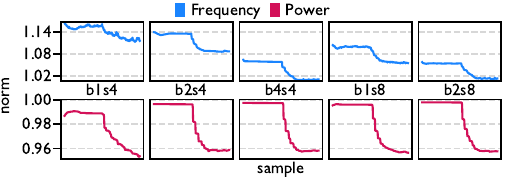}
	\caption{
		Measured frequency and power for different configurations of GPU-Red normalized to the minimum and maximum respectively of all configurations.
		A rolling window extracts the 5th and 95th quantile of 2000 samples for frequency and power respectively.
		Tuning begins halfway.
	}%
	\label{fig:GPU_Red_freq_power_window}
\end{figure}

\begin{figure}[!t]
	\centering
	\includegraphics[width=\columnwidth]{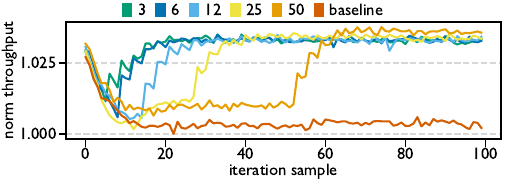}
	\caption{
		Different warm-up periods swept.
		Baseline is the default settings for GPU-Realloc with no power capping.
	}%
	\label{fig:GPU_Realloc_warmup_sweep}
\end{figure}

\begin{figure}[!t]
	\centering
	\includegraphics[width=\columnwidth]{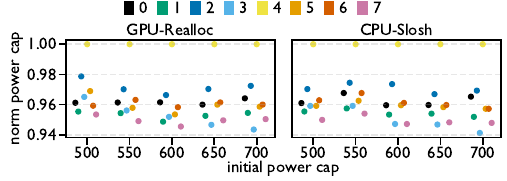}
	\caption{
		Final power caps set for different scenarios and initial power caps.
		Default settings from Table~\ref{tab:sens_study} are used.
	}%
	\label{fig:power_dist}
\end{figure}

\subsection{Sensitivity Study}%
\label{sec:sens_study}

In this section, we sweep values in Table~\ref{tab:sens_study} to determine their impact on power and throughput.

\begin{figure*}[!t]
	\includegraphics[width=\textwidth]{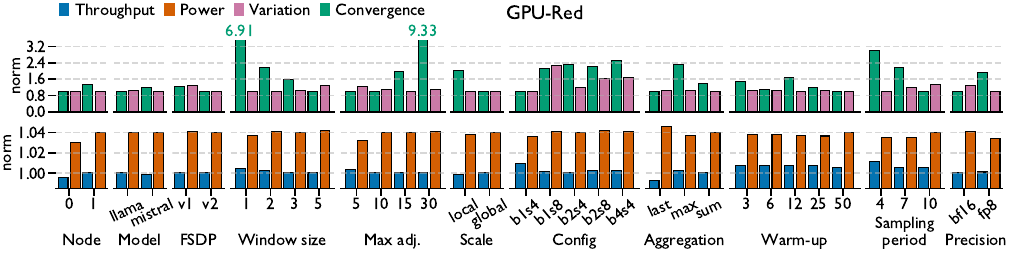}
	\caption{
		Sensitivity study of knobs in Table~\ref{tab:sens_study}.
		A higher value is better (e.g., less variation has a larger bar value).
		The rolling average of power from Figure~\ref{fig:GPU_Red_freq_power_window} is used for power reduction, and convergence as the number of samples between 99.5\% of max power, and 100.5\% of min power.
		Raw power samples as in Figure~\ref{fig:total_power_and_freq_all_use_cases} after convergence are used to measure variation in power ($CV = \sigma/\mu$).
		The mean of the last five values prior and post adjustment are used to calculate throughput improvement.
		Exceptions are warm-up and sampling period which are normalized to a baseline with no power-capping.
	}%
	\label{fig:GPU_Red_sensitivity_study}
\end{figure*}

\minisection{GPU-Red.}
Figure~\ref{fig:GPU_Red_freq_power_window} shows a power reduction of $4\%$ is achieved across all configurations.
While the average frequency varies across configurations, they all decrease proportionally with power.
This demonstrates that \name is present to the same degree across different configurations.
Indeed, Figure~\ref{fig:GPU_Red_sensitivity_study} demonstrates consistent power savings with maintained throughput across nearly all knobs.
However, there are a few exceptions.
Node 0 has more stragglers than node 1, illustrated in Figure~\ref{fig:lead_val}, and cannot reduce power on as many leaders as node 1.
Additionally, some knobs with worse convergence (e.g., max adj. 5) achieved worse power reduction.
In this case, power reduction was limited by the length of the experiment.
Given more iterations, their power reduction would match other knobs.

\begin{figure}[!t]
	\includegraphics[width=\columnwidth]{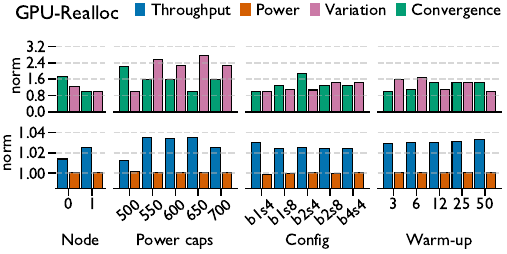}
	\caption{
		Power and throughput metrics are the same as Figure~\ref{fig:GPU_Red_sensitivity_study}.
		Convergence is measured as the samples needed for throughput to reach 99.5\% of peak.
		Variation in throughput is measured after the convergence point ($CV = \sigma/\mu$).
	}%
	\label{fig:GPU_Realloc_sensitivity_study}
\end{figure}

\minisection{GPU-Realloc.}
A throughput improvement between $2.5 \%$ and $3.5 \%$ is achieved across nearly all knobs in Figure~\ref{fig:GPU_Realloc_sensitivity_study}.
However, we observe lower throughput improvement on node 0 due to having fewer leaders to take power from, similar to worse power improvement in GPU-Red.
Additionally, a power cap of 500W has lower throughput improvement.
This power cap has significantly worse variation than other configurations, indicating volatility when running at some power caps.
Finally, Figure~\ref{fig:GPU_Realloc_warmup_sweep} illustrates that throughput converges to similar values regardless of warm-up length, confirming that power adjustments should be made immediately.

\begin{figure}[!t]
	\includegraphics[width=\columnwidth]{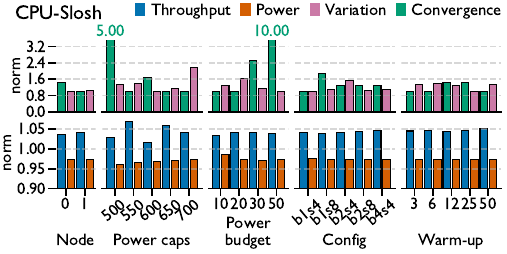}
	\caption{
		Metrics are the same as Figure~\ref{fig:GPU_Realloc_sensitivity_study}.
	}%
	\label{fig:CPU_Slosh_sensitivity_study}
\end{figure}

\minisection{CPU-Slosh.}
Figure~\ref{fig:CPU_Slosh_sensitivity_study} shows a consistent throughput improvement of $4 \%$ across all knobs, up to $6 \%$ for a power cap of 550W.
Additionally, we observe that after a power budget of 20W, no more power is consumed by the GPUs.
This is the case where the system has reached peak throughput, and is reducing power to maintain it like GPU-Red.

\minisection{Takeaway.}
We observed minor differences across different knobs in Figures~\ref{fig:GPU_Red_sensitivity_study},~\ref{fig:GPU_Realloc_sensitivity_study}, and~\ref{fig:CPU_Slosh_sensitivity_study}.
The most influential variable was the initial power cap used.
Despite this, the final power-caps set for different initial power caps have a very similar distribution as shown in Figure~\ref{fig:power_dist}.
This demonstrates that after a converged power distribution has been determined, it can be re-used for different frameworks, models, power-caps, and other knobs in Table~\ref{tab:sens_study}.
Re-usability is critical for a datacenter with dynamic node-level power caps, and diverse workloads.

\subsection{Mixture of Experts (MoE) Training}%
\label{sec:moe}

MoE models replace the standard MLP layer with multiple experts.
Tokens are routed to the corresponding expert(s), which improves inference throughput and lowers training time by only activating a subset of all experts.
To train these models, expert parallelism is commonly used which assigns unique experts to each GPU~\cite{outrageously_large_moe,switch_transformers}.
This introduces a cost to route tokens to each expert via all-to-all communication.
Unlike all-gather and reduce-scatter, all-to-all communication in MoE usually does not overlap with computation.
Since the number of tokens sent to each expert varies, workload imbalance can manifest in communication and computation.

To determine the impact of \name{} on MoE training, we use Primus, AMD's recommended training platform with torchtitan~\cite{torchtitan} as a backend to train DeepSeek V3 16B~\cite{deepseek} with eight-way expert parallelism.
This setting pads GEMMs, resulting in balanced MoE weight computation.
While there is experimental research targeting non-padded GEMMs, it is not yet supported on our platform~\cite{moe_grouped_gemm, sonic_moe}.

\begin{figure}[!t]
	\centering
	\begin{subfigure}[b]{0.489\textwidth}
		\includegraphics[width=\columnwidth]{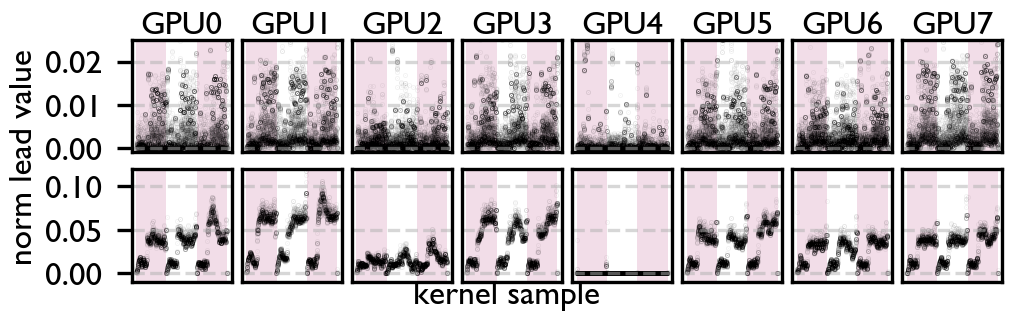}
		\caption{
			Lead values for DeepSeek (top row) and Llama (bottom row) pre-adjustment using the same metrics as Figure~\ref{fig:lead_val}.
			Large lead spikes occur frequently for DeepSeek.
			Zooming into 2\% and 10\% of the maximum spike, we see stragglers are the same for DeepSeek and Llama (i.e., GPU4).
			Since all-to-all communication is not overlapped for DeepSeek, GPUs are synchronized every layer, resulting in very small lead values relative to Llama.
		}%
		\label{fig:deepseek_llama_straggler_per_gpu}
	\end{subfigure}
	\begin{subfigure}[b]{0.489\textwidth}
		\includegraphics[width=\columnwidth]{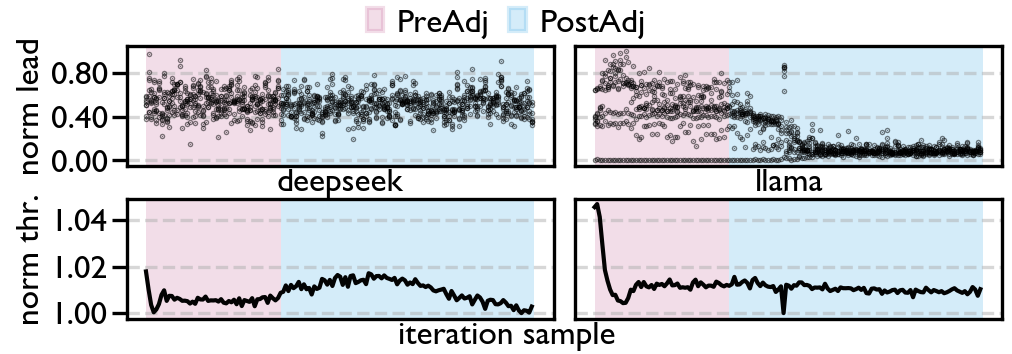}
		\caption{
			Aggregated lead values and throughput using the same metrics as ~\ref{fig:lead_and_throughput_all_use_cases}.
			The large spikes in lead value from DeepSeek inflate the aggregate summed lead value, despite most lead values being small relative to Llama as shown in Figure~\ref{fig:deepseek_llama_straggler_per_gpu}.
		}%
		\label{fig:deepseek_llama_lead_and_throughput}
	\end{subfigure}
	\begin{subfigure}[b]{0.489\textwidth}
		\includegraphics[width=\columnwidth]{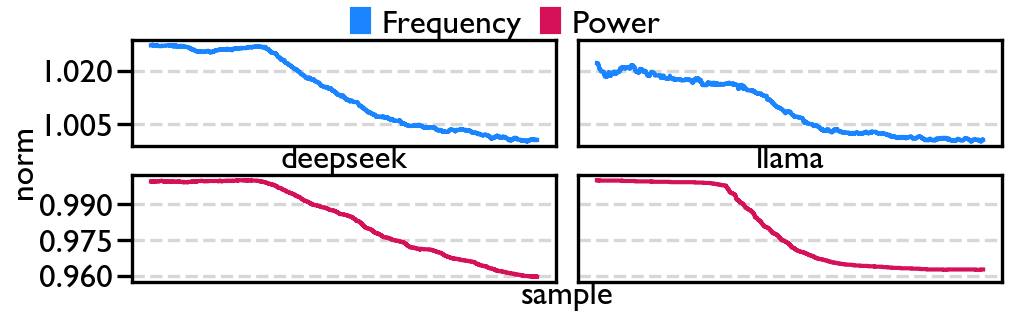}
		\caption{
			Measured frequency for DeepSeek and Llama, using the same metrics as Figure~\ref{fig:GPU_Red_freq_power_window}.
			Tuning begins one third of the way.
			Dense and MoE training exhibit similar power and frequency characteristics despite different communication collectives and model architectures.
		}%
		\label{fig:deepseek_llama_average_power_freq}
	\end{subfigure}
	\caption{
		Comparison of Llama 3 8B (b2s4) dense training and DeepSeek v3 16B (b8s4) MoE training using GPU-Red with defaults in Table~\ref{tab:sens_study}.
	}%
	\label{fig:deepseek_vs_llama_comparison}
\end{figure}

We compare dense to MoE training using Llama and DeepSeek in Figure~\ref{fig:deepseek_vs_llama_comparison}.
DeepSeek has more variation than Llama in lead values, throughput, and power.
This is because the expert parallel implementation does not overlap all-to-all collectives, causing frequent GPU synchronization every layer, compared to synchronization at the end of an iteration in dense training.
Since synchronization is more frequent, the lead resets every layer, causing smaller lead values relative to Llama in general. We also observe occasional high latency communication spikes, manifesting as very large lead values as shown in the scale of Figure~\ref{fig:deepseek_llama_straggler_per_gpu}.
Both small lead values and large spikes make it more difficult to classify leaders and stragglers.
Despite this, our algorithm still succeeds in finding a stable power distribution, matching the power savings of dense training as shown in Figure~\ref{fig:deepseek_llama_average_power_freq}.

\subsection{Detection Frequency and Overhead}

Over a three-month period, we tuned \name twice, achieving power savings of 3.5\% and 4\%.
Therefore, we recommend detection frequency at the week or month granularity to maximize benefits and minimize overhead.
Regarding the overhead, as shown in Figure~\ref{fig:GPU_Realloc_warmup_sweep}, power is stabilized within as few as 20 samples, and we sample one out of every ten iterations.
Given that dumping and processing one sample takes roughly 4 seconds, approximately 80 seconds of detection and mitigation are needed to reach a stable power distribution.

\section{Discussion}
\label{sec:discussion}

\subsection{Cost Savings}

Here we estimate the cost saving if our solution were deployed to the full datacenter.
Llama3 405B was trained on 16K GPUs, with each node applying tensor parallelism, which lies well within our assumption of a uniform workload~\cite{grattafiori2024llama3herdmodels} in a datacenter.
While FSDP might not be running on every node of a cluster for small-scale workloads, any balanced workload, regardless of C3, can benefit from aligning frequencies.
Even for imbalanced workloads like MoE, benefits have been shown in Section~\ref{sec:moe}.
Therefore, we argue that \name is applicable to a variety of datacenter scenarios.

Both OpenAI and Meta recently announced a partnership with AMD to deploy 6 gigawatts of AMD GPUs each~\cite{amd_openai_partnership, amd_meta_partnership}.
Google reports a Power Usage Effectiveness (PUE), a ratio of total datacenter energy to computing equipment energy, of $1.09$ across their own datacenters, with an industry average of $1.56$~\cite{google_pue_efficiency}.
GPU power is approximately 50\% of the provisioned power, and power usage for training and inference is reported to average 75\% of TDP~\cite{LLMPowerManMicrosoft}. Given the average price of electricity as of August 2025 is $\$0.14$~\cite{eia_kwh_cost},
a 4\% power saving could translate to over \$70 million saved annually for one customer.
\begin{align*}
	 & 6 \text{GW} / 1.56 \times 50\% \times 75\%                                                \\
	 & \times (24 \times 365)\text{h} \times 0.14\,\$/\text{kWh} \times 4\% \approx \$70\text{M}
\end{align*}

\subsection{Synergy with AI Trends}
\minisection{Lower Precision.}
As AI training and inference in general move towards lower precision, it is important to know what the impact of \name will be.
Figure~\ref{fig:GPU_Red_sensitivity_study} illustrates that \name is almost equally present for training in bf16 and fp8.
With more aggressive four-bit data, more studies are needed to understand how \name impacts.

\minisection{Inference Applicability}
Given the fundamental nature of \name, we consider it as workload agnostic.
GPUs used for AI training and inference are often the same, and will experience the same thermally induced straggling.
AI inference also utilizes C3~\cite{ficco}, meaning it can suffer from \name.

\minisection{Reliability Effects.}
Specifications exist which provide guidance on safely exceeding TDP, for certain magnitudes and certain timescales already~\cite{whitney2019ocp}.

\minisection{Multi-tenancy.}
\name{} describes thermal imbalance and variation from C3 in a balanced workload like FSDP training, and is more difficult to address when there is imbalance across GPUs as in multi-tenancy.
However, multi-tenancy often uses resource partitioning to allow for deterministic performance (e.g., split a GPU into training and inference partitions with CU masking~\cite{rapidserve, krisp}).
In such imbalanced cases, inter-GPU synchronization still exists, causing \name.
If partitions on a GPU are using the same resources, this could introduce variation in addition to thermal imbalance.
Even with such variation, stable and repetitive computation phases would be still observable in order to meet service level objectives.

\minisection{Accelerators.}
Since accelerators can have more deterministic performance than GPUs, we expect thermal/frequency effects to dominate the remaining variation and correlate with straggling at least as strongly as on GPUs.
However, accelerators typically use DMA for inter-device communication, whose behavior is more complex and warrants further study.

\subsection{Production Deployment}%
\label{sec:production}

While our current solution relies on users having administrator privileges to tune power caps, multi-tenant clusters usually cannot grant users these privileges.
However, there are other possible solutions to mitigate \name{} in both multi-tenant and private clusters.
For rapid, online frequency tuning, a firmware solution triggered by user-level application hints could synchronize frequencies between GPUs using GPU telemetry instead of user provided lead values; either through the CPU or between GPUs.
This online solution may require additional hardware for telemetry and synchronization.
For infrequent, offline tuning, a hook could run a stress test like our benchmark to calibrate GPUs intermittently (e.g., when a node is idle) since the ideal power caps remain relatively constant as shown in Figure~\ref{fig:power_dist}.
This offline solution could be deployed today without additional hardware, but may not be as efficient as online tuning.
Section~\ref{sec:thermal} and Figure~\ref{fig:temp_freq_corr} show that temperature and frequency, though correlated, are not perfectly matched, indicating potential GPU-inherent variation (e.g., induced by manufacture).
That said, prior work shows that GPU placement within a node can also affect thermal imbalance~\cite{sego}, suggesting variations in manufacturing and cooling can jointly cause straggling.

\subsection{Limitation}
Theoretically, \name applies to all systems with multiple devices in a node, where per-device DVFS is equipped.
We leave broader validation for future work, including AI accelerators, GPUs from other vendors, and beyond.
Also, this work is limited to a single node, and it is worthy to expand our solution at the cluster level and understand the impact for large-scale AI training.
Furthermore, given the prevalence of LLM inference with KV cache in industry frameworks such as vLLM~\cite{vllm_paper}, it is extremely beneficial to incorporate our solutions into such frameworks as default optimizations.

\subsection{Related Works}
\minisection{Straggler handling.}
Both datacenter-level and node-level solutions exist.
Datacenter-level solution identifies that the major source of stragglers is workload, such as uneven pipeline stage partitioning and imbalance in sequence lengths across batches, rather than hardware or software~\cite{straggler_rootcause_ai_paper}.
Node-level solutions propose optimized communication collectives to better hide the straggler idle time to improve resource utilization~\cite{stragglAR}.

\minisection{Energy saving.}
A lot of prior works focus on reducing the energy consumption without impacting the performance significantly.
Primary energy bottlenecks includes the uneven model pipelining and hardware straggling~\cite{energy_bloat_paper}.
Example solutions are power oversubscription, frequency locking and power capping, and fine-grained DVFS~\cite{LLMPowerManMicrosoft, micro-serve_paper, fine_grained_dvfs}.

\minisection{C3 mitigation}
Multiple techniques has been proposed to mitigate the slowdown due to C3.
Knowing the potential of C3 to improve performance, architecture support has been extended to support more efficient and finer-grained overlap~\cite{dma_gpu_paper_2}.
To further bridge the gap from theoretical performance, efforts have been made to design better communication collectives~\cite{ConCCL}.

DMA engines free compute resources from communication kernels, lowering the runtime variation of compute kernels during C3.
Since DMA does not eliminate the coupling between thermal imbalance and C3, \name{} can still exist.
However, the quantitative impact on lead values is complicated, which are determined by both the overlap and runtime of all preceding kernels.
That said, solving \name{} still provides benefits, since frequency differences across GPUs determine power and performance, as stated in Insights~\ref{ins:perf-model}~and~\ref{ins:power-model}.

\section{Conclusion}
\label{sec:conclusion}

In this paper, we identify the \name effect for a single-node multi-GPU system, which reveals how thermally induced straggling couples with C3 to impact performance variation and inefficiency.
We build performance and power models to understand the gains of solving \name.
We further propose a lightweight solution to detect and mitigate \name in real hardware and software systems, using only about 200 lines of PyTorch code.
Our solution can improve the performance and power by $6\%$ and $4\%$, respectively.

\section{Acknowledgment}
We thank all reviewers for their valuable feedback.
This work was sponsored by the Funding for Academic Research Program (gift funding) under the AMD University Program.
Access to GPUs was provided by the AMD University Program AI \& HPC Cluster and the AMD Developer Cloud.

AMD, AMD Instinct, AMD EPYC, and combinations thereof are trademarks of Advanced Micro Devices, Inc.
Other product names used in this publication are for identification purposes only and may be trademarks of their respective companies.

\balance{}
\bibliographystyle{IEEEtranS}
\bibliography{refs}

\end{document}